\documentclass[twocolumn,showpacs,preprintnumbers,amsmath,amssymb]{revtex4}
\usepackage{graphicx}
\usepackage{dcolumn}
\usepackage{bm}
\usepackage{color} 
\def\v#1{\mbox{\boldmath $#1$}}

\begin{document}
\title{Universal direction in thermoosmosis of a near-critical binary  fluid mixture
}

\author{Shunsuke Yabunaka}\email{yabunaka123@gmail.com}
\affiliation{Advanced Science Research Center, Japan Atomic Energy Agency, Tokai, 319-1195, Japan}

\author{Youhei Fujitani}
 \email{youhei@appi.keio.ac.jp}
\affiliation{School of Fundamental Science and Technology,
Keio University, 
Yokohama 223-8522, Japan}

\date{\today}

\begin{abstract}
{We consider thermoosmosis of 
a near-critical binary fluid mixture, lying in the one-phase region,
through a capillary tube in the presence of 
preferential adsorption of one component. 
The critical composition is assumed in the two reservoirs linked by the tube.
With coarse-grained approach,  we  evaluate the flow field induced by the thermal force density.
We predict a universal property;
if the mixture is near the upper (lower) consolute point,
the flow direction is the same as (opposite to) the direction of the temperature gradient, irrespective of which component 
is adsorbed onto the wall.}
\end{abstract}
\maketitle
A temperature gradient in a fluid along the confining surface can generate
a force density, parallel to the gradient, due to inhomogeneity in a fluid region 
near the surface.  This {{\it thermal force density} \cite{piazza,Wurger}} causes the {slip velocity}
across the region and drive the fluid in the bulk.
This bulk mass flow is called thermoosmosis,  {which does not involve
 gravity responsible for the Reyliegh-Benard convection.}  
Momentum transfer via the local slip can also induce
thermophoresis --- migration of a colloidal {object} 
in a fluid under a temperature gradient. 
Understanding these phenomena involves a fundamental problem in nonequilibrium physics and 
will lead to effective manipulations on lab-on-a-chip processes \cite{piazza,marbach,chen,
diffphore, diffdrop}.  
\par
Derjaguin and Sidorenkov (DS) observed {thermoosmosis of water through porous glasses} \cite{derja3}.  
{Applying the continuum theory and Onsager's reciprocity,}
they proposed a formula expressing the thermal force density 
in terms of the local excess enthalpy for {a one-component fluid} \cite{derja,derja2,anders}. 
According to this formula, the direction of
the flow is the same as (opposite to) that of 
the temperature gradient if the excess enthalpy density is negative (positive) {everywhere near the wall}.  
This is expected naively {by considering that} the flow in this direction tends to eliminate the temperature gradient  
by carrying the fluid with lower (higher) enthalpy to the region with higher (lower) temperature.
However, the local excess enthalpy is not easy to access experimentally and
is numerically evaluated only on the basis of simplified microscopic models \cite{fu,ganti}.  Besides, 
well-definedness of microscopic expression of excess enthalpy is questioned especially near the surface \cite{ganti,Anzini}. 
Therefore, it remains difficult to incorporate detailed microscopic interactions theoretically and even predicting the flow direction is often challenging \cite{piazza}.
In Ref.~\cite{ganti},  the authors {propose an extension of} DS's formula for multicomponent fluids
{within continuum description, while questioning its validity} in a microscopic slip layer.   
\par
\begin{figure}
\includegraphics[width=8cm]{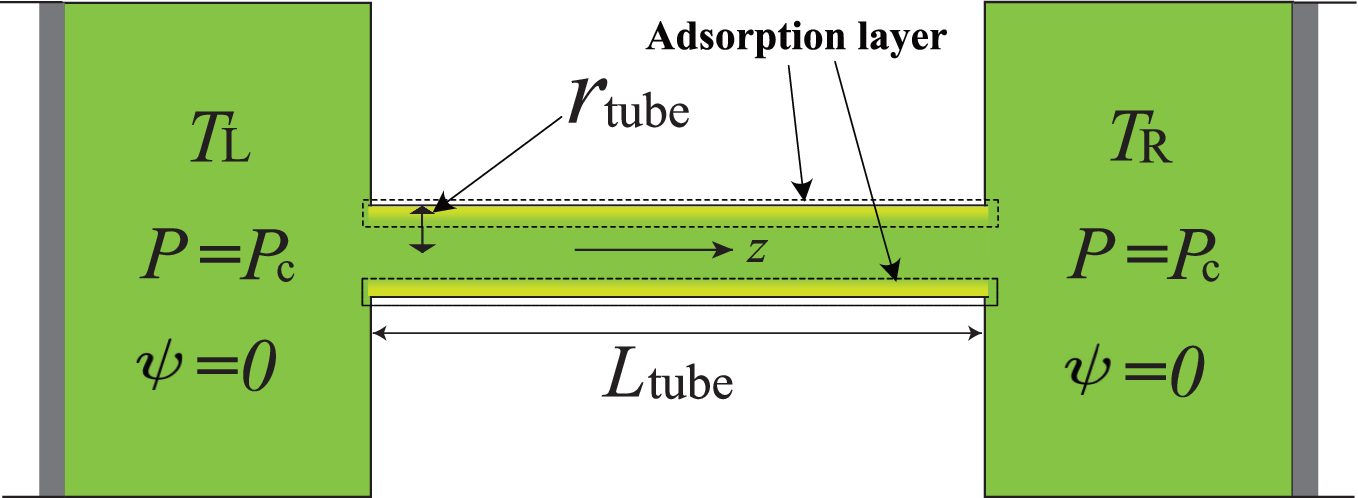}
\caption{Schematic of our setting.
A mixture is filled in the container composed of two reservoirs
and a capillary tube \textcolor{magenta}{between them} {with} the radius $r_{\rm tube}$ and the length $L_{\rm tube}$. 
The $z$ axis is taken along the tube and is directed to the right reservoir.  
One component, drawn in yellow, is preferentially adsorbed
onto the tube's wall. Thick walls represent pistons.
The pressure $P$ and the order parameter {$\psi$}
in the reservoirs are always set equal to {their values at the critical point},  
$P_{\rm c}$ and zero, respectively.
The temperatures in the left and right reservoirs, denoted by $T_{\rm L}$ and $T_{\rm R}$, respectively, 
are equal to $T^{({\rm ref})}$ in the reference state.
}
\label{fig:tube}
\end{figure}

 
Thermoosmosis has not been studied in relation to  critical phenomena, to our best knowledge.
In this Letter, we study thermoosmosis of a  binary fluid mixture near the demixing critical point
through a capillary tube linking two large reservoirs (Fig.~\ref{fig:tube}). 
The mixture is assumed to lie in the one-phase region throughout inside the container, and is
simply referred to as a mixture in the following. We assume preferential adsorption (PA) of one component on the tube's wall 
due to short-range interactions.
The adsorption layer, enriched by the preferred component, was first observed in Ref.~\cite{beysens1982},
and are studied theoretically  \cite{binder, diehl86, diehl97, fisher-degennes, RJ,law}.
{The layer becomes much thicker than} the molecular sizes near the critical point.
Therefore, we can 
evaluate the universal properties of thermal force density with  continuum description, 
avoiding difficulties associated with microscopic approach discussed in the last paragraph.
When a temperature difference is imposed between the reservoirs, as shown later,
the thermal force density is generated in the adsorption layer to {cause thermoosmosis.}
{We} predict a universal property;
the flow direction is the same as (opposite to) {the direction of} the temperature gradient 
in thermoosmosis of a mixture near the upper (lower) consolute point, irrespective of which component 
is adsorbed on the wall.
\par
The tube is assumed to be a cylinder having the
radius $r_{\rm tube}$ and length $L_{\rm tube}$ (Fig.~\ref{fig:tube}). 
We write $\rho_{\rm a}$ ($\rho_{\rm b}$) for the mass density of a mixture component named a (b),
{defining $\rho$ as
$\rho_{\rm a}+\rho_{\rm b}$ and $\varphi$ as $\rho_{\rm a}-\rho_{\rm b}$.
In general, the scalar pressure and the temperature are respectively denoted by $P$ and $T$.
In the container}, we first prepare an equilibrium one-phase state of the mixture, which we call a reference state. 
{This state is specified by $P^{({\rm ref})}=P_{\rm c}$, $\varphi^{({\rm ref})}=\varphi_{\rm c}$ and $T^{({\rm ref})}(\approx T_{\rm c})$. 
The superscript $^{({\rm ref})}$ (subscript $_{\rm c}$) indicates a value in the reservoirs in the reference state
(a value at the critical point).}
{The order parameter $\psi$, defined as $\varphi-\varphi_{\rm c}$, vanishes in the reference state.
The value of $T$ in the right (left) reservoir is denoted by $T_{\rm R}$ ($T_{\rm L}$),
which equals $T^{({\rm ref})}(\approx T_{\rm c})$ in the reference state.}  
Next, we slightly change {$T_{\rm R}$ and $T_{\rm L}$ from $T^{({\rm ref})}$
to make $\delta T$ nonzero, where
$\delta T$ is defined as $T_{\rm R}-T_{\rm L}$,} {while keeping
$P$ and $\psi$ in the reservoirs at $P_c$ and} zero, respectively (Fig.~\ref{fig:tube}).
The mixture is approximately incompressible under usual experimental conditions.  
Hence, we assume $\rho=\rho_{\rm c}$
 throughout inside the container in this Letter.   
\par
{We write $\tau$ for the reduced temperature $(T-T_{\rm c})/T_{\rm c}$.}
The demixing critical point can be an upper consolute (UC) point or a lower consolute (LC) point \cite{sciam,kaji,tsori}.
Near a UC (LC) point, {$\tau$ is positive (negative)  {in} the one-phase region.} 
We here {roughly} explain our key idea by
using the Landau model, whose free-energy density is given by a quadratic function of $\psi^2$.
The density includes a term $a\tau\psi^2$, where 
$a$ is a positive (negative) constant near the UC (LC) point.  {By 
operating $-T^2\partial_T T^{-1}$ on the free-energy density, we} {find 
this term to contribute $-a\psi^2$ to 
the internal-energy density, which is 
negative (positive) in the adsorption layer of a mixture near a UC (LC) point. }
Hence, assuming that the contribution is dominant in the excess enthalpy density, 
which is mentioned in the second paragraph,  
we {can conjecture} that
the thermoosmotic direction of a mixture near a UC (LC) point
is the same as (opposite to) the direction of the temperature gradient, irrespective of which component is preferred by the tube's wall. 
\par
\begin{figure}
\includegraphics[width=6cm]{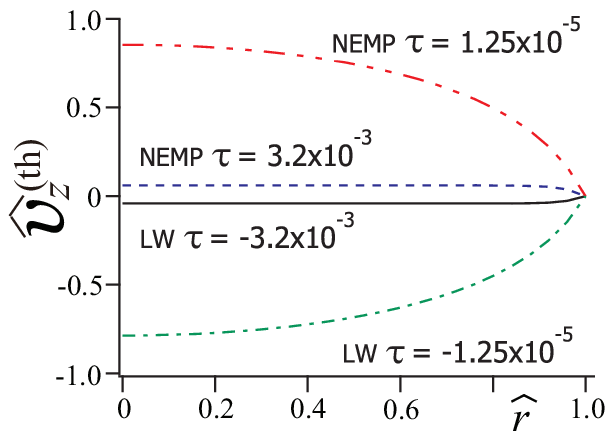}
\caption{The $z$ component of the dimensionless velocity field in thermoosmosis, ${\hat v}_z^{({\rm th})}({\hat r})$, 
is plotted against the dimensionless radial distance ${\hat r}$ for a mixture of NEMP (LW) near the UC (LC) point.  
The reduced temperature $\tau$ of the reference state,
indicated in the figure, is positive (negative) in the one-phase region near the UC (LC) point.
The other parameter values are the same as used in Fig.~\ref{fig:inte}.}
\label{fig:velocityprofile}
\end{figure}

{To examine the conjecture stated above,} we apply the hydrodynamic formulation under inhomogeneous temperature \cite{dvw,gonn} and
the renormalized local functional theory (RLFT) \cite{fisher-degennes,rlft}.
{We} consider a weak, stationary, and laminar flow in the tube, which is
so thin and long that effects of the tube edges 
on the flow are negligible. 
The no-slip boundary condition is imposed on the tube's wall,
which is {impermeable} and adiabatic.
On a tube's cross section, we write $r$ for the radial distance from the center
and define a dimensionless radial distance ${\hat r}$ as $r/r_{\rm tube}$.  The $z$ axis is taken as in Fig.~\ref{fig:tube}. {We {consider} a mixture of 
2,6-lutidine and water (LW) {\cite{mirz}} near the LC point and a mixture of nitroethane and 3-methylpentane (NEMP)  {\cite{iwan}}
near the UC point. }
Before describing the details, {we show   
the velocity profile under $\delta T>0$ in Fig.~\ref{fig:velocityprofile}},
{where} the flow direction is the same as (opposite to) the direction of the temperature gradient 
in a mixture near the UC (LC) point  {and} the flow rate
is larger in magnitude as the critical temperature is approached.
\par
We assume that the free-energy density in the bulk of a mixture, $f_{\rm bulk}$,
is a function of $\rho_{\rm a}$, $\rho_{\rm b}$, the quadratic form of their gradients, and $T$;
$f_{\rm bulk}$ is coarse-grained up to the local correlation length of the order-parameter fluctuations, $\xi$.
Hydrodynamics is applicable to flow whose typical length is locally larger than $\xi$. 
The chemical potential $\mu_n$ conjugate to $\rho_n$ is given by
\begin{equation}
\mu_n=\frac{\partial f_{\rm bulk}}{\partial\rho_n}-T\nabla\cdot\left[ \frac{1}{T}\frac{\partial f_{\rm bulk}}{\partial\left(\nabla
\rho_n\right)}\right]\ .\label{eqn:chempot}\end{equation}
The reversible part of the pressure tensor, denoted by ${\mathsf \Pi}$, is symmetric and is given by
\begin{equation}
{\mathsf \Pi}=P {\mathsf 1}+ 
\sum_{n={\rm a,b}} \left(\nabla \rho_n\right) \frac{\partial f_{\rm bulk}}{\partial\left(\nabla\rho_n\right)}
\ ,\label{eqn:Pidef}\end{equation}
where ${\mathsf 1}$ is the identity tensor of order two
and $P$ equals the negative of the grand-potential density.  
Defining $\mu_\pm$ as $(\mu_{\rm a}\pm \mu_{\rm b})/2$, we have
$P=\mu_+\rho+\mu_-\varphi-f_{\rm bulk}$.
Equations (\ref{eqn:chempot}) and (\ref{eqn:Pidef}) are derived for a one-component fluid in Ref.~\cite{dvw}
and are applied to a binary fluid mixture \cite{gonn}, {assuming that the coefficient of the gradient term is linear with respect to $T$ in the free-energy density.} A complete set of the hydrodynamic equations are shown in Sect.~II of Ref.~\cite{companion}.
In its Appendix A, Eqs.~(\ref{eqn:chempot}) and (\ref{eqn:Pidef}) are derived without the assumption.  
The internal energy and entropy per unit volume are denoted by $u$ and $s$, respectively. 
The partial entropy and enthalpy per unit mass of the component $n$ are
denoted by ${\bar s}_n$ and ${\bar H}_n$, respectively.
We define ${\bar s}_-$ and ${\bar H}_-$ as $({\bar s}_{\rm a}-{\bar s}_{\rm b})/2$ and
$({\bar H}_{\rm a}-{\bar H}_{\rm b})/2 = \mu_-+T{\bar s}_-$, respectively.  
\par
We linearize the dynamics with respect to $\delta T$. {Difference between the reservoirs is indicated by
$\delta$}, such as $\delta T$.  The Gibbs-Duhem (GD) relation gives
\begin{equation} 
0=\rho^{({\rm ref})}\delta\mu_+ +\varphi^{({\rm ref})} \delta\mu_- +s^{({\rm ref})} \delta T
\ ,\label{eqn:gd}\end{equation}
where $\delta\mu_-$ equals $-{\bar s}_-^{({\rm ref})} \delta T$ on our assumption $\delta P=\delta\rho=\delta\varphi=0$.
In the tube, the mass conservation gives $\nabla\cdot {\v v}=
\partial_z v_z=0$ with ${\v v}$ denoting the velocity field.
There, the momentum conservation gives
\begin{equation}
2\nabla\cdot\left(\eta_{\rm s}{\mathsf E}
\right)=\nabla\cdot {\mathsf \Pi}\ ,
\label{eqn:stok}
\end{equation}
{where ${\mathsf E}$ denotes the rate-of-strain tensor with
${\mathsf E}_{ij}=(\partial _i v_j +\partial _j v_i)/2$ in the Cartesian coordinates}. 
The shear viscosity, denoted by $\eta_{\rm s}$, generally depends on the position via its dependence on $\xi$. 
Equations (\ref{eqn:chempot}) and (\ref{eqn:Pidef}) yield  
\begin{equation}
\nabla\cdot {\mathsf\Pi}
=s\nabla T + \sum_{n={\rm a,b}} \left( \rho_n\nabla\mu_n + \frac{\nabla T}{T}\cdot 
\frac{\partial f_{\rm bulk}}{\partial \left(\nabla \rho_n\right)} \nabla\rho_n\right) ,\label{eqn:nablaPi}
\end{equation}
which can be regarded as an extended GD relation.
Combined with the irreversible terms, this extended GD relation
guarantees positive entropy production in bulk and
the Onsager's reciprocity for osmotic fluxes through the tube.  
These points, which justify {using} Eqs. (1)-(5) in our derivation of thermal force density, are shown in Ref.~\cite{dvw} and Appendix B of Ref.~\cite{companion}, respectively. 
\par
{We add the superscript $^{({\rm th})}$ to a quantity in the tube in the linear regime of thermoosmosis we consider.
The thermal force density, denoted by $\sigma_z^{({\rm th})}$,  is given by the 
$z$-component of $-\nabla\cdot {\mathsf\Pi}$ on this condition, where
Eq.~(\ref{eqn:nablaPi}) has only $z$ component.}
The {conservation equations for energy and mass densities and their}  boundary conditions are satisfied if $\mu_\pm$ and $T$
are linear functions of $z$ and homogeneous on a tube's cross section.
See Sect.~IIC of Ref.~\cite{companion} for details.
Using Eqs.~(\ref{eqn:gd})--(\ref{eqn:nablaPi}), we find $\sigma_z^{({\rm th})}$ 
{to be dependent only on $r$ and to be given by} 
$-{\delta T}/({T^{(\rm ref)}} L_{\rm tube})$ multiplied by
\begin{equation}
u(r)+P(r)-u^{({\rm ref})}-P^{({\rm ref})}-\bar{H}_{-}^{({\rm ref})}\psi(r)\ ,
\label{eqn:thermal-force-density}\end{equation}
where $u(r)$, $P(r)$, and $\psi(r)$ are evaluated in the tube in the reference state 
and thus $P(r)$ equals ${\mathsf \Pi}_{zz}(r)$. 
{This formula is an extension of DS's formula to two-component fluids, since the first four terms {of Eq.~(\ref{eqn:thermal-force-density})}
can be regarded as} the excess enthalpy density in DS's formula 
for {a one-component fluid}.  
Our procedure to derive the formula for the thermal force density
via an extended GD relation could {also be} applied to 
{any soft material} described with a free-energy functional.  
We compare our derivation of the 
thermal force density with the corresponding part in Ref.~\cite{ganti} as follows. 
Because the sum of the last three terms of Eq.~(\ref{eqn:thermal-force-density}) equals
{$-\rho_{\rm a}{\bar H}_{\rm a}^{({\rm ref})}-\rho_{\rm b}{\bar H}_{\rm b}^{({\rm ref})}$}, 
our formula for {$-\sigma^{({\rm th})}_z$, given by the product of 
${\delta T}/({T^{(\rm ref)}} L_{\rm tube})$ and Eq.~(\ref{eqn:thermal-force-density}), formally coincides with
the right-hand side (RHS)} of Eq.~(5) of Ref.~\cite{ganti}, where they interpret the RHS as {$-\sigma^{({\rm th})}_z$}.
However, its {left-hand side (LHS)}, {$\partial_z {\mathsf \Pi}_{zz}$ in our notation},  
is not equal to {$-\sigma^{({\rm th})}_z$} in general, 
since {$\partial_x {\mathsf \Pi}_{xz}+\partial_y {\mathsf \Pi}_{yz}$} does not vanish 
in the presence of PA. Here, $x$ and $y$ are orthogonal coordinates
on the tube's cross section.
In Ref.~\cite{ganti}, this sum {$\partial_x {\mathsf \Pi}_{xz}+\partial_y {\mathsf \Pi}_{yz}$} 
is also missing in {the LHS of} Eq.~(2), which the authors employ
as an extended GD relation in deriving their Eq.~(5). {The sum should be included in the extended GD relation 
for deriving the formula of the thermal force density properly. }
\par
We have $v_z=0$ at $r=r_{\rm tube}$ owing to the no-slip condition and  $\partial_{r}v_{z}=0$ at $r=0$ 
owing to the axissymmetry and smoothness of $v_z$.  Thus,
the $z$ component of Eq.~(\ref{eqn:stok}) gives
 \begin{equation}
 v_{z}^{({\rm th})}(r)=
\int_{r}^{r_{\rm tube}}dr_1\ \int_{0}^{r_1}dr_2\ \frac{r_2 \sigma_z^{({\rm th})}(r_2)}{ r_1\eta_{\rm s}(r_1)}
\ ,\label{eqn:vzprofile2}\end{equation}
where $\eta_{\rm s}$ is evaluated in the reference state and depends on the radial distance. 
In the absence of PA, Eq.~(\ref{eqn:thermal-force-density}) vanishes and
thermoosmosis does not occur.  \par
{The} correlation length and, therefore, the effects of critical fluctuations become spatially inhomogeneous   inside the adsorption layer \cite{RJ}. To describe these effects, 
we introduce a coarse-grained free-energy functional as follows.  {We write $k_{\rm B}$ for the Boltzmann constant, 
and} use the conventional notation for the critical exponents --- $\beta, \gamma, \nu,$ and $\eta$.  
The (hyper)scaling relations give $2\beta+\gamma=3\nu$ and
$\gamma=\nu(2-\eta)$; we adopt $\nu= 0.630$ and $\eta= 0.0364$ \cite{peli}.
A mixture with $\psi=0$ has $\xi=\xi_0|\tau|^{-\nu}$, where $\xi_0$ is 
a material constant.  The functional consists of two terms. One is given by
an area integral of $-h \varphi$ over the wall, representing the wall-component interactions. 
The constant $h$, called the surface field, vanishes in the absence of PA \cite{binder,diehl86,diehl97}.
The other is given by the volume integral of $f_{\rm bulk}$.
We neglect the coupling between $\rho$ and $\psi$, {considering the mixture's incompressibility}. 
{Under the chemical potentials $\mu_n^{({\rm ref})}$,
the} grand-potential density in the bulk is 
{${f_{\rm bulk}}-\rho_{\rm a}{\mu}_{\rm a}^{({\rm ref})}-\rho_{\rm b}{\mu}_{\rm b}^{({\rm ref})}$}.  
According to the RLFT {\cite{fisher-degennes, rlft}}, 
its $\psi$-dependent part is $k_{\rm B}T$
multiplied by the sum of
\begin{equation}
\frac{1}{2} C_1\xi_0^{-2}\omega^{\gamma-1}|\tau|\psi^2 +
\frac{1}{12} C_1C_2\xi_0^{-2} \omega^{\gamma-2\beta} \psi^4 \label{eqn:rlftinteg}
\end{equation} 
and the square gradient term,  $C_{1}\omega^{-\eta\nu} \left\vert\nabla\psi\right\vert^2/2$. 
See Sect.~IIIC of Ref.~\cite{companion} for the rest part.
Here, $C_1$ and $C_2$ are material constants satisfying
$C_{2}=3u^{*}C_{1}\xi_{0}$, where $u^*$ is the scaled coupling constant at the Wilson-Fisher fixed point
and equals $2\pi^{2}/9$ at the one loop order.  The local ``distance'' from the critical point
is represented by $\omega\equiv (\xi_0/\xi)^{1/\nu}$, which leads to $\omega=|\tau|$ if $\psi$ vanishes.
The self-consistent condition,
$\omega={\left|\tau\right|}+C_{2}\omega^{1-2\beta}\psi^{2}$, locally determines how $\xi$ depends on $\tau$ and $\psi$.
As in Ref.~\cite{rlft},
we can obtain $\psi$ in the reference state by minimizing the $\psi$-dependent part of the total 
grand potential. 
This is equivalent to solving Eq.~(\ref{eqn:chempot}) with {$\mu_n=\mu_n^{({\rm ref})}$}
and $T=T^{({\rm ref})}$ under {the boundary condition involving the surface field.} 
\par
Below, we introduce critical scalings  in terms of $r_{\rm tube}$. We
define $T_*$ so that $\xi$ becomes $r_{\rm tube}$ for $\psi=0$ at $T=T_*$ in the one-phase region,
define $\tau_*$ as $|T_*-T_{\rm c}|/T_{\rm c}$, and introduce a scaled {reduced-temperature}
${\hat \tau}\equiv \tau/\tau_*$.
A characteristic order parameter $\psi_*$ is defined so that
$\xi$ becomes  $r_{\rm tube}$  for $\psi=\psi_*$ 
at $T=T_{\rm c}$, and a characteristic chemical potential
$\mu_*$ is defined as {$k_{\rm B}{{T_*}}/(3u^*r_{\rm tube}^3 \psi_*)$.
The} scaled surface field ${\hat h}$ is defined as $h {T_*} /\left(T\mu_*r_{\rm tube}\right)$.
 {We define
the dimensionless equilibrium profile ${\hat \psi}({\hat r})$ as $\psi(r)/\psi_*$}. 
A dimensionless function ${\hat f}({\hat \psi})$ is defined as Eq.~(\ref{eqn:rlftinteg}) divided by
$T_*/(\mu_*\psi_*T)$ and is given by
 \begin{equation}
{\hat f}({\hat\psi}){=\frac{1}{2}{\hat\omega}^{\gamma-1}
\left|{\hat \tau}\right|{\hat\psi}^2 +\frac{1}{12}{\hat\omega}^{\gamma-2\beta}
{\hat\psi}^4}\ .\label{eqn:prefmm0}\end{equation}
The first term on the RHS above originates from $a\tau\psi^2$
in the Landau model, or more precisely, the corresponding term in the bare $\psi^4$ model.
{In the reference state}, ${\hat \psi}({\hat r})$
is determined only by $|{\hat \tau}|$ and ${\hat h}$.
The scaled thermal force density ${\hat\sigma}_z^{({\rm th})}({\hat r})$, defined as
${\sigma}_z^{({\rm th})}(r) {\tau_{*}}{T_{*}}L_{\rm tube}/(\mu_{*}\psi_{*}\delta T)$, is found to be
\begin{equation}
{\tau_*}\left({\hat f}+\frac{|\partial_{\hat r}{\hat \psi}|^2}{2{\hat\omega}^{\eta\nu}}\right)+
{\frac{T^{({\rm ref})}}{T_{\rm c}}}\left(\frac{\partial {\hat f}}{\partial{\hat \tau}}+
{\frac{\partial{\hat\omega}^{-\eta\nu}}{\partial{\hat \tau}}}\frac{|\partial_{\hat r} {\hat\psi}|^2}{2}\right)
\ ,\label{eqn:thermal-force-density0}
\end{equation}
{which is evaluated in the reference state.  Here},
${\hat \omega}\equiv \omega/\tau_*$ is regarded as a function of ${\hat \tau}$ and ${\hat \psi}$ via the self-consistent condition.
See Sect.~III D {of Ref.~\cite{companion}} for the details.
Hereafter, $\tau$ ($\hat{\tau}$) represents the (scaled) reduced temperature in the reference state.
\par
\begin{figure}
\includegraphics[width=6cm]{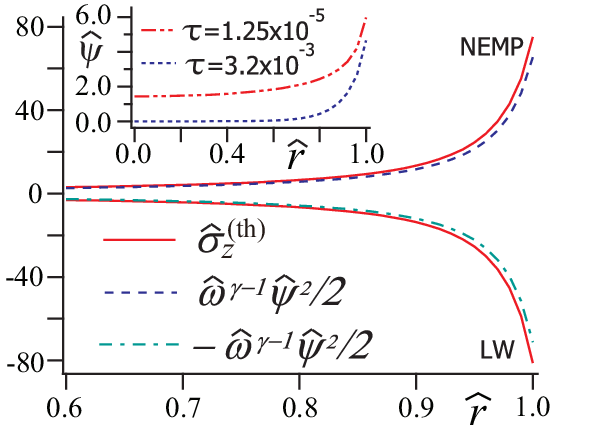}
\caption{Scaled thermal force density Eq.~(\ref{eqn:thermal-force-density0}) (red solid curves)
and its dominant term are plotted against the dimensionless radial distance ${\hat r}$
for $|\tau|=1.25\times 10^{-5}$ and $h=0.1 \ $cm$^3$/s$^2$. 
The results for a mixture of NEMP (LW) near the UC (LC) point are shown in the upper (lower) half of the panel.
The blue dashed curve (green dash-dot curve) represents the dominant term
${(-)}{\hat \omega}^{\gamma-1}{\hat\psi}^2/2$.
(Inset) For $h=0.1 \ $cm$^3$/s$^2$,
the dimensionless equilibrium profile ${\hat\psi}({\hat r})$ in a mixture of NEMP 
is plotted against ${\hat r}$ at $\tau=1.25\times 10^{-5}$ {and $3.2\times 10^{-3}$}.}
\label{fig:inte}
\end{figure}
We study the profile of  ${\hat \sigma_z}^{({\rm th})}{({\hat r})}$ 
given by Eq. (\ref{eqn:thermal-force-density0}) to determine the direction of thermoosmosis. Here
we set $r_{\rm tube}$ equal to $0.1\ \mu$m and use the same values of the material constants
as in Ref.~\cite{pipe}.  The {parameter} values are {summarized} in Table I of Ref.~\cite{companion}.
In particular, for a mixture of LW (NEMP), we find $T_{\rm c}=307\ (300)$K,
$\xi_0=0.20\ (0.23)\ $nm, and $\tau_*=5.12\times 10^{-5}\ (6.49\times 10^{-5})$
from the experimental data of Refs.~\cite{mirz, iwan}, 
and set ${\hat h}$ to $73.0$ ($66.6$), which amounts to $h=0.1 \ $cm$^3$/s$^2$.
{Rough estimation of $h$ is given in Sect.~VI of Ref.~\cite{yabufuji}}. 
The red solid curves
in Fig.~\ref{fig:inte} {indicate ${\hat\sigma}_z^{({\rm th})}({\hat r})$ given by
Eq.~(\ref{eqn:thermal-force-density0}).}  The sum in its second parentheses is denoted by ${\hat\sigma}_z^{({\rm th}2)}$.
We numerically 
{find ${\hat\sigma}_z^{({\rm th}2)}  \approx {\hat\sigma}_z^{({\rm th})}$, which}
is reasonable since $\tau_*\ll 1$ 
and $T^{({\rm ref})}/T_{\rm c}\approx 1$.  
{Notably, ${\hat\sigma}_z^{({\rm th}2)}$} 
is determined only by scaled quantities ${\hat\tau}$ and $|{\hat h}|$ in the framework of 
the RLFT.  As can be seen from Eq.~(\ref{eqn:prefmm0}),
$\partial {\hat f}/(\partial {\hat \tau})$ 
contains $\pm{\hat \omega}^{\gamma-1}{\hat\psi}^2/2$,
 {where the same sign  as $\tau$ is taken.}  
This term is dominant in Eq.~(\ref{eqn:thermal-force-density0}), 
according to our numerical results in Fig.~\ref{fig:inte}.
See {Sections} IVA and IVC of Ref.~\cite{companion} for more details. The signs of $\sigma_z^{({\rm th})}$ and ${\hat \sigma_z}^{({\rm th})}$ 
are the same when $\delta T$ is positive.  Thus, 
${\hat \sigma_z}^{({\rm th})}{({\hat r})}>0\ (<0)$ for $0\le {\hat r}\le 1$ means that the direction of 
 the flow is the same as (opposite to) that of the temperature gradient. {Notably, Eq.~(\ref{eqn:thermal-force-density0})  does not contain $\bar{H}_{-}^{({\rm ref})}$,
and remains the same if the sign of $h$ is changed, which indicates that the direction of thermoosmosis is independent of which component 
is preferentially adsorbed on the wall.}
{The curves in the inset of Fig.~\ref{fig:inte}, representing ${\hat \psi}({\hat r})$ of a mixture of NEMP in the reference state, 
rise near the wall because of $h>0$ and show that the adsorption layer is thicker at the smaller
value of $\tau$.  For $\tau=1.25\times 10^{-5}\ (3.2\times 10^{-3})$, $\xi/r_{\rm tube}$ is equal to $0.032\ (0.038)$ 
at ${\hat r}=1$ and to $0.47\ (0.086)$ at ${\hat r}=0$.}
\par
{Finally we study the velocity field  $v_z^{({\rm th})}({\hat r})$ given by Eq.~(\ref{eqn:vzprofile2}).}  
The viscosity in Eq.~(\ref{eqn:vzprofile2}) weakly diverges {near} the critical point   
\cite{halhohsig,ohta,bergmold}. 
In Appendix E of Ref.~\cite{pipe}, 
we obtain the viscosity as a function of $|\tau|$ 
and $\psi$ from the results of Refs.~\cite{bhatt,tsai} and find the value of $\eta_*$, 
{which is defined as the viscosity's singular part at $\psi=0$ and $T=T_*$},
from the data of Refs.~\cite{gratt,stein,iwan,leis}. 
{In Fig.~\ref{fig:velocityprofile}, we plot the dimensionless velocity field ${\hat v}_z^{({\rm th})}({\hat r})$ defined as $v_z^{({\rm th})}(r)T_*\tau_*\eta_*L_{\rm tube}/(\mu_*\psi_* r_{\rm tube}^2\delta T)$.}
At $|\tau|=3.2\times 10^{-3}$ in this figure,
{${\hat v}_z^{({\rm th})}({\hat r})$} {appears} to change only in the region of {${\hat r}>0.8$} and thus the velocity appears to slip across this 
region.   {This is reasonable since the adsorption layer, where the thermal force is nonvanishing, localizes
near the wall} at $|\tau|=3.2\times 10^{-3}$, as shown in the inset of Fig. \ref{fig:inte}.  
The value at the flat portion of the black solid (blue dashed) curve is $-0.042\ (0.061)$,
which means that the slip velocity across the adsorption layer is 
$-7.09\ (38.2)\ (\mu{\rm m})^2/($s$\cdot$K$)$ multiplied by $\delta T/L_{\rm tube}$.
{These values are comparable in magnitude  
with a typical value} measured for thermophoretic mobility {\cite{piazza,Jiang,Braun,maeda}}. 
{For example, if we set {$|\delta T|=100\ {\rm mK} \ll |T^{(\rm ref)}-T_{\rm c}|\approx 1\ $K} and $L_{\rm tube}=10\ \mu$m,
the slip velocity is approximately {$0.1\ \mu$m$/$s}, which would be
measured experimentally}.
At $|\tau|=1.25\times 10^{-5}$, the slip is not clear in Fig.~\ref{fig:velocityprofile}, because
the thermal force density decreases gradually as $\hat{r}$ decreases  as shown in Fig.~\ref{fig:inte}. 
\par
To conclude, {we predict that,} for {any binary fluid mixture in the one-phase region 
near the upper (lower) consolute point},  
the {direction in thermoosmotic flow} is the same as (opposite to) that of the temperature gradient, 
irrespective of which component is adsorbed onto
the tube's wall, if the critical composition is assumed in 
the reservoirs.  
In the companion paper \cite{companion}, we consider the Onsager coefficients linking
general thermodynamic forces and fluxes through a tube.  Our coarse-grained approach could be
applied to thermoosmosis of
polymer solutions and {polyelectrolytes} \cite{onukibook} and thermophoresis of colloidal particles driven 
by the thermal force density near the surface, for example, 
{with} interactions relevant for mesoscopic structures {taken into account}.\\

We acknowledge Takeaki Araki, Masato Itami, Yusuke T. Maeda, Kouki Nakata, Yuki Uematsu,  
and Natsuhiko Yoshinaga for careful reading the manuscript and giving comments.  
S. Y. was supported by Grant-in-Aid for Young Scientists  (18K13516).

\end{document}